# AppleCiDEr. II. SpectraNet—A Spectroscopic Neural Network Classifier for Transients Demonstrated on ZTF Follow-up Data

Maojie Xu[1,2], Argyro Sasli[2,3], Alexandra Junell[2,3], Felipe Fontinele Nunes[2,3], Yu-Jing Qin[4], Christoffer Fremling[5,6], Sam Rose[7], Theophile Jegou Du Laz[2,4], Benny Border[2,3], Antoine Le Calloch[2,3], Sushant Sharma Chaudhary[2,3], Hailey Markoff[2,3], Avyukt Raghuvanshi[2,3], Nabeel Rehemtulla[8,9,10], Jesper Sollerman[11], Yashvi Sharma[5], Niharika Sravan[2,12], Judy Adler[13], Tracy X. Chen[13], Richard Dekany[5], Reed Riddle[5], Mansi M. Kasliwal[6], Matthew J. Graham[2,14], and Michael W. Coughlin[2,3]

[1] Department of Computer Science & Engineering, University of Minnesota, Minneapolis, MN 55455, USA; xu000810@umn.edu
[2] NSF Institute on Accelerated AI Algorithms for Data-Driven Discovery (A3D3), USA
[3] School of Physics and Astronomy, University of Minnesota, Minneapolis, MN 55455, USA
[4] Division of Physics, Mathematics and Astronomy, California Institute of Technology, 1200 E California Blvd., Pasadena, CA 91125, USA
[5] Caltech Optical Observatories, California Institute of Technology, Pasadena, CA 91125, USA
[6] Division of Physics, Mathematics and Astronomy, California Institute of Technology, Pasadena, CA 91125, USA
[7] Cahill Center for Astrophysics, California Institute of Technology, MC 249-17, 1216 E California Boulevard, Pasadena, CA 91125, USA
[8] Department of Physics and Astronomy, Northwestern University, 2145 Sheridan Road, Evanston, IL 60208, USA
[9] Center for Interdisciplinary Exploration and Research in Astrophysics (CIERA), 1800 Sherman Ave., Evanston, IL 60201, USA
[10] NSF-Simons AI Institute for the Sky (SkAI), 172 E. Chestnut St., Chicago, IL 60611, USA
[11] Department of Astronomy, Stockholm University, 10691 Stockholm, Sweden
[12] Department of Physics, Drexel University, Philadelphia, PA 19104, USA
[13] IPAC, California Institute of Technology, 1200 E. California Blvd, Pasadena, CA 91125, USA
[14] Cahill Center for Astrophysics, California Institute of Technology, Pasadena, CA 91125, USA
*Received 2025 October 9; revised 2026 July 29; accepted 2026 August 13; published 2026 September 15*

## Abstract

Time-domain surveys such as the Zwicky Transient Facility have opened a new frontier in the discovery and characterization of transients. While photometric light curves provide broad temporal coverage, spectroscopic observations remain crucial for physical interpretation and source classification. However, existing spectral analysis methods—often reliant on template fitting or parametric models—are limited in their ability to capture the complex and evolving spectra characteristic of such sources, which are sometimes only available at low resolution. In this work, we introduce `SpectraNet`, a deep convolutional neural network designed to learn robust representations of optical spectra from transients. Our model combines multi-scale convolution kernels and pooling operations to extract features from preprocessed spectra in a hierarchical and interpretable manner. We train and validate `SpectraNet` on low-resolution time-series spectra obtained from the Spectral Energy Distribution Machine and other instruments, demonstrating better performance in classification compared to other known pipelines.

## 1. Introduction

Time-domain surveys, such as the Zwicky Transient Facility (ZTF; Bellm et al. 2019; Graham et al. 2019; Masci et al. 2019; Dekany et al. 2020), have revolutionized our understanding of the sky by delivering high-cadence, wide-field photometry, enabling the detection and follow-up observation of a diverse range of transient events. Supernovae (SNe) are the most common class of transients, and demographic studies of their subtypes have a long history, with early systematic analyses already establishing key population trends (e.g., Drout et al. 2011; Modjaz et al. 2014). The large ZTF samples have continued this progress by expanding subtype statistics and enabling higher-cadence demographic studies (Perley et al. 2020; Dhawan et al. 2021).

ZTF has also played a crucial role for nuclear transients by building statistical samples of tidal disruption events (TDEs; van Velzen et al. 2021; Hammerstein et al. 2023) and identifying active galactic nuclei (AGNs; Frederick et al. 2019; Graham et al. 2020). Beyond these discoveries, the survey has led to studies for exotic fast transients (Andreoni et al. 2020; Ho et al. 2023) and compact binaries (Kupfer et al.

2019; Burdge et al. 2020). These highlight the broad discovery space accessible to wide-field high-cadence surveys.

Photometric data plays a crucial role in discovering and monitoring these events; however, many types of transients present degeneracies in their light-curves. Spectroscopy is a very powerful tool and can prevent misclassification and help distinguish between different underlying mechanisms and environments (e.g., Minkowski 1941; Craig Wheeler et al. 1993; Filippenko 1997; Branch et al. 2006; Osterbrock & Ferland 2006; Gal-Yam 2017; Zabludoff et al. 2021; Dong et al. 2022; González-Bañelos et al. 2025). In the time-domain context, spectra provide a dynamic fingerprint of evolving transients, with absorption and emission features evolving over timescales of hours to weeks, or even months and years. Spectroscopic follow-up is therefore essential for classification purposes, but also for providing physical insights about the transient.

To meet these requirements, a dedicated suite of spectroscopic facilities has been constructed. In the ZTF ecosystem, the Spectral Energy Distribution Machine (SEDM; Blagorodnova et al. 2018; Rigault et al. 2019; Kim et al. 2022) serves as the primary rapid-classification instrument and has been a workhorse for the Bright Transient Survey (Fremling et al. 2020; Perley et al. 2020; Rehemtulla et al. 2024). It is a low-resolution ($R \sim 100$) integral-field spectrograph installed on the 60 inch Palomar telescope, which enables timely identification and follow-up prioritization (Rehemtulla et al. 2025). However, many of the most compelling scientific results from ZTF discoveries rely on high signal-to-noise, high-resolution spectroscopy obtained with larger aperture facilities such as Keck (Oke et al. 1995; Faber et al. 2003), Gemini (Davies et al. 1997), LBT (Pogge et al. 2010), and others. These instruments provide detailed measurements of velocity structure, composition, ionization conditions, and host-galaxy environments that cannot be resolved with SEDM alone.

In addition to SEDM, an international network of instruments on a variety of telescopes also contributes to transient classification, including SDSS (Abdurro'uf et al. 2022), DESI (DESI Collaboration et al. 2024), and the Next Generation Palomar Spectrograph (NGPS; Kasliwal et al. 2024); the reader is referred to (Junell et al. 2026) for a more comprehensive list. Finally, software infrastructure also plays a crucial role in triggering follow-up observations and processing the resulting data, including `SkyPortal` (van der Walt et al. 2019; Coughlin et al. 2023), `Fritz`,[15] alternative management platforms such as YSE-PZ (Coulter et al. 2023), the GROWTH Marshal (Kasliwal et al. 2019), the Transient Name Server (TNS),[16] and public repositories such as `WISeREP` (Yaron & Gal-Yam 2012).

However, characterizing the collected data remains a major challenge. Template-based and parametric spectral classifiers, such as `SNID` (Blondin & Tonry 2007), `Superfit` (Howell et al. 2005), and `GELATO` (Harutyunyan et al. 2008), rely on comparisons to libraries of well-characterized supernova spectra. While effective for common classes and high-quality observations, these approaches can perform poorly for rare or unusual events, or when the continuum shape and emission-line behavior deviate from standard templates. Moreover, such models typically evaluate each spectral snapshot independently, neglecting valuable information encoded in the temporal evolution of spectral features (Modjaz et al. 2016; Yao et al. 2019).

More recently, deep learning-based models have been used to process spectroscopic data. One of the earliest deep-learning-based spectral classifiers was `DASH` (Muthukrishna et al. 2019), which uses a convolutional neural network to classify supernova spectra and estimate their phases. `SNIa-score` (Fremling et al. 2021) utilises BiLSTM and GRU layers to classify low-resolution Type Ia supernova (SN) spectra in real time. Similarly, `CCSNScore` (Sharma et al. 2025) employs a multi-input framework specifically designed to recognize and classify core-collapse SNe. `GalSpecNet` (Wu et al. 2024) employs a one-dimensional CNN for emission-line galaxy spectral classification, and its architecture was later adapted as the spectral encoder in the `AstroM`$^3$ pipeline (Rizhko & Bloom 2025). Another approach is the use of vision transformers (Junell et al. 2026; Strano Moraes et al. 2026) to examine spectra. Recently, (Fortino et al. 2026) introduced `ABC-SN`, a transformer-based spectral classifier for SNe that outperforms previous models, offering improved accuracy across ten SN subtypes. Some of the models discussed here specialize in specific transient classes, for example supernovae, or even limited subsets of supernovae (SN Ia or CC SNe). Motivated by these limitations, we introduce a novel deep-learning framework for spectral modeling in the time-domain regime. We present `SpectraNet`, a CNN-based architecture which uses multi-scale convolutional filters, and spectral pooling methods to learn robust features both at local and global wavelength scales. Rather than assuming that rare-class or cross-survey challenges are fully resolved, we treat them as important limitations that remain relevant when interpreting the results of the present ZTF-focused study.

The model is part of the broader `AppleCiDEr` framework (Junell et al. 2026), which integrates spectroscopic information with images, photometric light curves, and contextual metadata for multimodal transient classification. In the present paper, however, we restrict our evaluation to the spectroscopic component in the ZTF follow-up setting.

The inclusion of `SpectraNet` within the `AppleCiDEr` architecture improves overall classification performance relative to versions of the pipeline that do not incorporate a dedicated spectral model (see Junell et al. 2026 for a detailed comparison). In addition, SpectraNet serves as a real-time decision-making component for follow-up prioritization, enabling early identification of scientifically valuable targets. To this end, AppleCiDEr is currently deploying into SkyPortal (van der Walt et al. 2019; Coughlin et al. 2023) as a classification pipeline and will soon do so within our broker BOOM (Burst & Outburst Observations Monitor) and BABAMUL as a classification pipeline (Jegou du Laz et al. 2026).

[15] https://github.com/fritz-marshal/fritz
[16] https://www.wis-tns.org/

In this work, we present SpectraNet as a deep-learning classifier developed and evaluated for ZTF follow-up spectroscopy. While the architecture may be adaptable to spectra from other facilities, the validation presented in this paper is limited to the ZTF/SEDM-centered setting, and broader survey-agnostic use remains future work.

This paper is organized as follows. Section 2 provides details about the data set and the pre-processing steps needed to create consistent input from heterogeneous spectrographs. Section 3 presents the architecture and training strategy of the SpectraNet model, while Section 4 discusses our model choices and a comparison with other known networks. Section 5 outlines the performance of the adopted model. Finally, in Section 6, we discuss our results and future work.

## 2. Dataset Description

Our training data set was obtained from Fritz (see footnote 15), the science data platform currently used by the ZTF collaboration and built within the SkyPortal ecosystem, as well as from the GROWTH Marshal (Kasliwal et al. 2019), which served as the earlier platform for parts of the ZTF follow-up workflow, (van der Walt et al. 2019; Coughlin et al. 2023) supplemented by spectra from SDSS (Abdurro'uf et al. 2022), DESI (DESI Collaboration et al. 2024) and TNS. For label consistency and training stability, subclasses with very few instances were merged into their parent classes, for example "SN Ic-BL" to "SN Ic" and "SN Ia-pec" into "SN Ia."

We note that merging rare subclasses into broader parent categories is not without drawbacks. Peculiar subclasses, such as SN Ia-pec or SN Ic-BL, are often of high astrophysical interest and can carry important implications for cosmology and transient discovery. For instance, misidentifying peculiar SN Ia as normal SN Ia may introduce biases in Hubble-diagram analyses, while rare or extreme objects may be overlooked for detailed follow-up if not flagged appropriately.

Our choice reflects a deliberate trade-off between taxonomic granularity and robustness at early classification stages. Given the severe class imbalance inherent in spectroscopic transient samples, training on highly fragmented subclasses can lead to overfitting and unstable performance across surveys and instruments. In this work, we therefore adopt a coarse but astrophysically meaningful labeling scheme intended for rapid triage and follow-up triggering, rather than final subclass determination. Importantly, SpectraNet is designed to operate as part of a broader classification ecosystem: parent-class predictions can be refined by downstream, domain-specific classifiers or expert vetting once additional data become available. We further assess the impact of this label choice in Figure 5, where SpectraNet is evaluated on an expanded set of subclasses.

**Table 1**
Distribution of Transient Object Types per Unique Object ID in the Dataset

| Type | Count | Type | Count |
|---|---|---|---|
| SN Ia | 7328 | SN Ibn | 39 |
| SN II | 1559 | SN Ib/c | 34 |
| CV | 392 | SN Ia-pec | 26 |
| AGN | 343 | SN II-norm | 23 |
| SN IIn | 321 | SN Ia-02cx | 18 |
| SN Ic | 282 | SN Ia-CSM | 14 |
| SN Ib | 252 | Nova-like | 10 |
| SN IIb | 194 | SN Ia-03fg | 10 |
| SN IIP | 150 | SN Ib-pec | 4 |
| TDE | 136 | Galactic Nuclei | 3 |
| SN Ia-91T | 86 | SN IIL | 3 |
| SN Ic-BL | 83 | SN Icn | 3 |
| SN Ia-norm | 81 | SN Ien | 1 |
| SN Ia-91bg | 41 | SN II-pec | 1 |
| Novae | 43 | SN Ia-99aa | 1 |

To ensure the reliability of the training set, objects without a confident classification in the source labels (i.e., cases for which the cataloged or curator-assigned class confidence was below 50%) were excluded, as their true type could not be determined with sufficient certainty. This threshold refers to the confidence associated with the original data set labels and not to a posterior probability output by SpectraNet. The data set includes multiple spectra for a single transient when available. As described in Section 2.2, all data partitions in the five-fold cross-validation are performed at the object level, rather than at the individual-spectrum level. As a result, all spectra from a given transient are assigned exclusively to a single fold, ensuring that no object appears in more than one fold and preventing information leakage between folds. To further reduce redundancy, spectra that are extremely close in time and numerically identical (within a tolerance of $\epsilon = 10^{-6}$) are excluded from the data set. This strategy mitigates biases associated with slow spectral evolution and repeated observations, as discussed in Fortino et al. (2026). Table 1 presents the original distribution of transient types at the object level, where each unique object ID is counted once. Table 2 shows the corresponding distribution at the spectrum level, where multiple epochs from the same object are included when available. Table 3 summarizes the final class distribution after label merging.

**Table 2**
Distribution of Transient Object Types Over All Epochs in the Dataset

| Type | Count | Type | Count |
|---|---|---|---|
| SN Ia | 12120 | SN Ibn | 235 |
| SN II | 4375 | SN Ib/c | 92 |
| CV | 565 | SN Ia-pec | 113 |
| AGN | 478 | SN II-norm | 70 |
| SN IIn | 1057 | SN Ia-02cx | 70 |
| SN Ic | 1183 | SN Ia-CSM | 67 |
| SN Ib | 955 | Nova-like | 91 |
| SN IIb | 740 | SN Ia-03fg | 57 |
| SN IIP | 307 | SN Ib-pec | 15 |
| TDE | 752 | Galactic Nuclei | 3 |
| SN Ia-91T | 145 | SN IIL | 3 |
| SN Ic-BL | 463 | SN Icn | 73 |
| SN Ia-norm | 133 | SN Ien | 19 |
| SN Ia-91bg | 78 | SN II-pec | 24 |
| Novae | 71 | SN Ia-99aa | 1 |

**Table 3**
Final Distribution of Transient Classes After Label Merging and Filtering

| Merged Type | Count |
|---|---|
| SN Ia | 12,783 |
| SN II | 6748 |
| SN Ibc | 3582 |
| CV | 752 |
| TDE | 727 |
| AGN***** | 478 |

**Note.** The sample counts include multiple spectra from the same object when available. * SN Ia includes: SN Ia, SN Ia-91T, SN Ia-norm, SN Ia-91bg, SN Ia-pec, SN Ia-02cx, SN Ia-CSM, SN Ia-03fg, and SN Ia-99aa. ** SN II includes: SN II, SN IIn, SN IIP, SN IIb, SN II-norm, and SN II-pec. *** SN Ibc includes: SN Ib, SN Ic, SN Ic-BL, SN Ibn, SN Ib/c, SN Icn, SN Ib-pec, and SN Ien. **** CV includes: CV, Nova-like, Novae. ***** AGN includes: AGN and Galactic Nuclei.

## 2.1. Spectral Preprocessing and Calibration

Each spectrum undergoes a multi-stage preprocessing pipeline to ensure physical consistency, numerical stability, and uniformity of input dimensions for downstream modeling. The process combines astrophysical calibrations with numerical data preparation and is applied identically across spectra from all instruments. This calibration is particularly important because our data set includes spectra obtained from a variety of instruments (e.g.,SEDM, DESI, SDSS, etc.).

*Wavelength filtering and validation.* We restrict all spectra to a shared wavelength interval of [3500 Å, 8500 Å]. To accommodate gaps due to detector edges or sky subtraction artifacts, we fill any remaining missing flux values with zeros. This masking strategy allows the model to learn to ignore missing regions without introducing artificial interpolation artifacts.

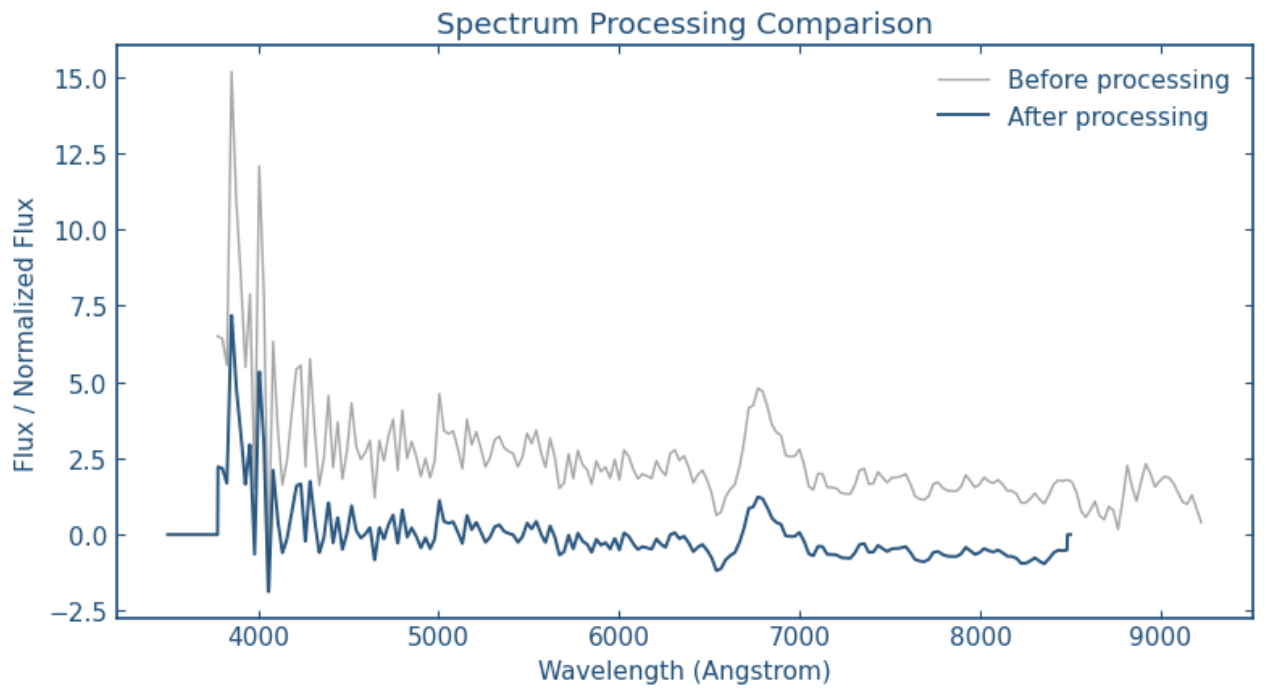

**Figure 1.** An example of spectral representation before (gray) and after (blue) preprocessing for the source "ZTF25acgtjkh."

*Interpolation to fixed grid.* All spectra are interpolated to a uniform grid of 4096 wavelength points using linear interpolation. This step ensures a consistent input shape for the model, and preserves the large-scale structure and line features of the original data. The interpolation grid is fixed across the data set to promote shared spatial alignment.

*Flux normalization.* Finally, to account for variability in source brightness and observational conditions, we normalize the flux values across each spectrum. We evaluated both min–max scaling to the [0, 1] range and Z-score normalization; we adopt the latter as the default, as it preserves the sign and dynamic range of absorption and emission features. For each spectrum with flux values $f$, the Z-score transformation is applied as

$$f_{\mathrm{norm}} = \frac{f - \mu_f}{\sigma_f},$$

where $\mu_f$ and $\sigma_f$ denote the mean and standard deviation of that spectrum's flux values. This standardization enhances numerical stability during training and mitigates the influence of amplitude variation due to distance or exposure time. An example of spectral representation before and after preprocessing is given in Figure 1.

Finally, we note that all spectra are used in the observed frame; no redshift correction is applied, as reliable redshifts are available for only a minority of the sample.

## 2.2. Data Splitting Strategy

For the transient type classification task, we adopt a five-fold cross-validation strategy to ensure unbiased evaluation and optimal utilization of the labeled samples. The data set is divided into five mutually exclusive and approximately equal-sized folds. In each iteration, four folds (80%) are used for training and the remaining fold (20%) is used for validation. Stratified sampling is employed to preserve balanced class distributions across all folds, thereby ensuring proportional representation of each transient class.

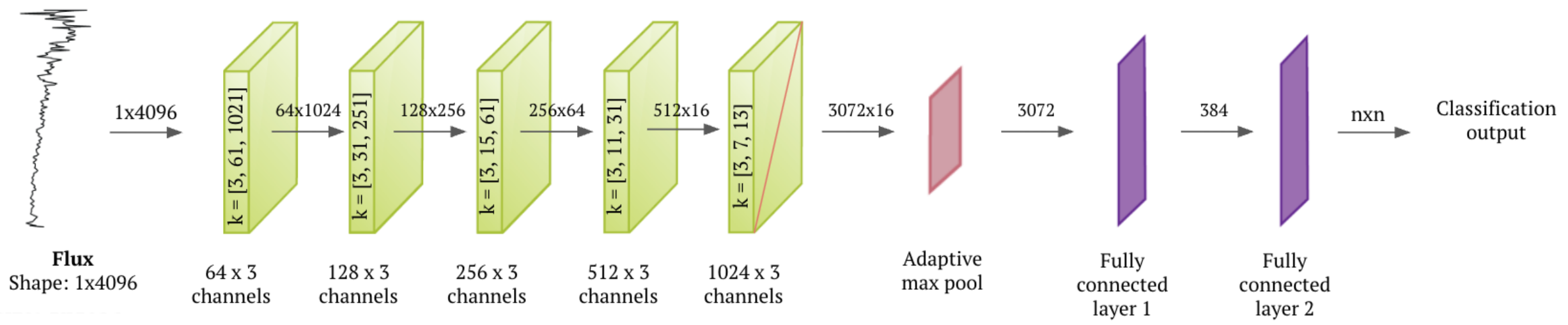

**Figure 2.** Schematic architecture of `SpectraNet`. The green block represents the *SpectraBlock*, and the red line in the last block indicates that no pooling and downsampling are implemented. The three kernel sizes correspond to small-, medium-, and large-scale 1D convolutions, as shown in Figure 3.

We use the macro-averaged F1 score evaluated on the validation set within each fold to guide model selection.[17] We adopt macro F1 as the primary selection criterion because it gives equal weight to all classes and is therefore more appropriate for our imbalanced classification setting. For completeness, we report accuracy, top-3 accuracy, macro-averaged F1, and macro AUROC as evaluation metrics.

The final model performance is obtained by averaging the results over all five cross-validation folds, in order to obtain a robust estimate of the generalization ability.

### 2.3. Data Augmentation

To improve robustness to missing or interpolated regions, we introduce an additional augmentation step: random masking. At every training epoch, a randomly selected region is masked either on the left (3500–6000 Å) or on the right (6000–8500 Å) side of each input spectrum. Moreover, the number of masked data points is randomly chosen to vary from 0 to 2048. This augmentation simulates realistic sensor or processing artifacts and encourages the model to learn more robust and distributed spectral representations. The masked region changes dynamically with each epoch to avoid overfitting to a fixed missing pattern.

## 3. Model Architecture and Training Strategy

Most existing 1D convolutional neural networks (1D CNNs) for spectral classification adopt single-scale convolutional kernels (Sharma et al. 2020; Chen 2021). Although such designs are simple and efficient, they often fail to capture the diversity of structures in astrophysical spectra, which may contain both narrow emission lines and broad continuum components.

Inspired by multi-branch convolutional architectures in image classification, such as the Inception module (Szegedy et al. 2015), we design each `SpectraBlock` with three parallel 1D convolutional paths, using small (e.g., 3), medium (e.g., 31), and large (e.g., 1021) kernel sizes $k$, for example, see Figures 2 and 3. Figure 2 provides a simplified diagram of the `SpectraNet` architecture, and Figure 3 illustrates the structure of each `SpectraBlock`. This design enables the model to simultaneously capture narrow spectral features, such as emission and absorption lines, and broader structures, such as continuum variations or blended features.

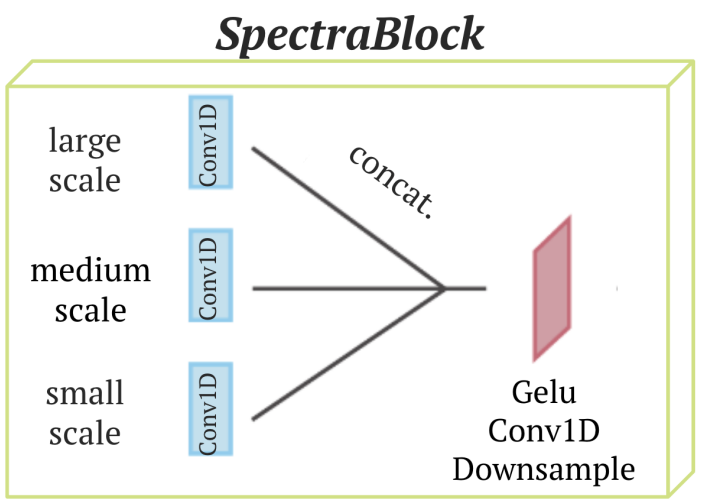

**Figure 3.** Schematic architecture of the `SpectraBlock`.

The outputs of the three 1D CNNs are concatenated along the channel dimension to form a high-dimensional multi-scale feature map of size $k \times C$ (e.g., $3 \times 64 = 192$ channels). Without further compression, this rapidly increases the number of parameters and computational burden, and may lead to overfitting—especially when stacking multiple such blocks.

To address this, we adopt a $1 \times 1$ convolution to project the concatenated feature map back to the $C$ output channels. This lightweight projection layer, also known as a pointwise convolution, serves two purposes: (1) it compresses the dimensionality to reduce computation, and (2) it enables the network to learn meaningful combinations of multi-scale features across channels. Crucially, the $1 \times 1$ operation maintains the same spectral resolution and does not increase the receptive field, making it ideal for channel-wise transformation.

The use of $1 \times 1$ convolution has been widely adopted in modern CNN architectures. Inception modules use it for both dimensionality reduction and multi-scale fusion (Szegedy et al. 2015), ResNet applies it for matching residual dimensions (He et al. 2016), and MobileNet leverages it in depthwise separable convolutions for efficient feature transformation (Howard et al. 2017). Following this principle, we apply a $1 \times 1$ convolution at the end of each `SpectraBlock`

[17] Model selection is based on the validation macro-F1 performance within each fold, including criteria such as the number of epochs and regularization strength.

to project the output back to a unified channel dimension before feeding it into the next stage.

After concatenation of the three 1D convolutions, a layer normalization is applied and a GELU activation function. Then, the features are downsampled to match the input shape of the next block, while the data points are reduced. However, no pooling or downsampling is applied in the final block. A schematic architecture of the `SpectraBlock` is given in Figure 3.

`SpectraNet` consists of five stacked *SpectraBlocks*, followed by an adaptive max pool and two fully connected layers. Figure 2 summarizes the overall network layout for the preprocessed $1 \times 4096$ spectral input.

### 3.1. Process Explanation

As illustrated in Figures 2 and 3, the input to `SpectraNet` is a preprocessed one-dimensional spectrum with a fixed length of 4096 wavelength elements. The spectrum is passed sequentially through five stacked *SpectraBlocks*. Within each block, three parallel one-dimensional convolutional branches use different kernel sizes to extract spectral features at small, intermediate, and large wavelength scales. The resulting feature maps are concatenated along the channel dimension and projected to the required output-channel dimension using a $1 \times 1$ convolution. Layer normalization and a GELU activation are then applied to the combined representation.

Between successive blocks, the spectral dimension is progressively reduced through downsampling while the number of feature channels increases. This allows the network to compress the wavelength axis while retaining increasingly abstract representations of narrow lines, broad spectral structures, and blended features. No pooling or downsampling is applied in the final *SpectraBlock*, thereby preserving the final-stage representation before global aggregation.

After the five *SpectraBlocks*, adaptive max pooling converts the variable feature map into a fixed-dimensional representation. This representation is then passed through two fully connected layers. The final layer produces one logit for each target class, and the logits are converted into class probabilities using a softmax function.

To mitigate the effects of class imbalance, particularly for rare transients such as TDEs, the model is trained using a class-balanced variant of focal loss (Lin et al. 2017). The class weights $\alpha$ are dynamically calculated using the effective-number-of-samples weighting scheme of Cui et al. (2019), reducing the influence of the substantial imbalance among the transient classes during optimization.

### 3.2. Training Configuration

We train the model using the AdamW (Loshchilov & Hutter 2019) optimizer with decoupled weight decay as a regularization mechanism. The initial learning rate is selected through Optuna hyperparameter optimization (Akiba et al. 2019). During the early training phase, we apply a linear warm-up schedule: the learning rate is linearly increased from zero to the target value over a predefined number of warmup epochs. After the warm-up phase, the learning rate is kept constant throughout the remaining training epochs.

The training process runs for a maximum of 50 epochs. To prevent overfitting, we adopt an early stopping strategy based on the macro-averaged F1 score evaluated on the validation set. Training is stopped early if the F1 score does not improve for 3 consecutive epochs (`patience`).

To further enhance the stability and generalization of the model, we incorporate an Exponential Moving Average (EMA; Tarvainen & Valpola 2017) of the model parameters. The EMA weights are applied during validation and for final model saving.

We also employ mixed-precision training using automatic mixed precision (AMP), which accelerates training and reduces GPU memory consumption without degrading model accuracy.

The final model selection is based on the highest macro-averaged F1 score achieved on the validation set during training.

## 4. Ablation Study

### 4.1. Design Choices: Kernel Size and Channel Expansion

We evaluate how convolution kernel size, output channel width, and normalization strategy affect model performance using five-fold cross-validation. All results are reported as mean $\pm$ standard deviation across the five folds.

*Kernel Size.* We compare single-scale and multi-scale kernel configurations to evaluate how receptive-field diversity affects spectral classification. As shown in Table 4, the proposed multi-kernel design achieves the best overall performance, with an accuracy of 82.54% $\pm$ 1.61%, a Top-3 accuracy of 97.01% $\pm$ 0.64%, a Macro F1 of 69.39% $\pm$ 3.13%, and a Macro AUC of 96.11% $\pm$ 0.49%. In comparison, the small-kernel configuration (3 across all stages) performs substantially worse, with an accuracy drop of 17.22 percentage points relative to the multi-kernel model. The medium-kernel setting (61, 31, 15, 11, 7 by stage) narrows this gap, but still trails the multi-kernel design by 4.80 points in accuracy. The large-kernel setting (1021, 251, 61, 31, 13 by stage) provides the strongest single-scale baseline and is only 1.38 points lower in accuracy, yet remains inferior to the multi-kernel model on all evaluation metrics. Overall, these results suggest that explicitly combining kernels with different receptive fields is more effective than using a single scale alone, even though this comes at the cost of reduced throughput (Table 4).

*Channel Width.* Table 5 compares different channel-width configurations to evaluate how model capacity affects spectral

**Table 4**
Study on Convolution Kernel Size Configurations

| Kernel Size Configuration | Accuracy (%) | Top-3 Acc (%) | Macro F1 (%) | Macro AUC (%) | Throughput (it/s) |
|---|---|---|---|---|---|
| Uniform Small ({3} across all stages) | 65.32 ± 1.50 | 92.13 ± 0.67 | 49.28 ± 2.19 | 88.14 ± 1.12 | **24.51** |
| Uniform Medium ({61, 31, 15, 11, 7} by stage) | 77.74 ± 0.61 | 96.14 ± 0.49 | 64.09 ± 0.86 | 94.21 ± 0.13 | 22.36 |
| Uniform Large ({1021, 251, 61, 31, 13} by stage) | 81.16 ± 0.54 | 96.39 ± 0.17 | 66.56 ± 1.67 | 95.57 ± 0.37 | 16.59 |
| Multi Kernel size | **82.54 ± 1.61** | **97.01 ± 0.64** | **69.39 ± 3.13** | **96.11 ± 0.49** | 11.48 |

**Note.** Each row specifies the kernel size strategy used across the network. All other settings are fixed. Bold values indicate the best result for each metric.

**Table 5**
Study on Channel Width Configurations

| Channel Widths (per stage) | Accuracy (%) | Top-3 Acc (%) | Macro F1 (%) | Macro AUC (%) | Throughput (it/s) |
|---|---|---|---|---|---|
| [1, 8, 16, 32, 64, 128] | 75.56 ± 1.53 | 95.07 ± 0.62 | 60.33 ± 1.84 | 93.34 ± 0.67 | **13.30** |
| [1, 16, 32, 64, 128, 256] | 77.33 ± 1.48 | 95.24 ± 0.58 | 61.43 ± 2.11 | 94.11 ± 0.71 | 13.25 |
| [1, 32, 64, 128, 256, 512] | 82.54 ± 1.61 | 97.01 ± 0.64 | 69.39 ± 3.13 | 96.11 ± 0.49 | 11.83 |
| [1, 64, 128, 256, 512, 1024] | **84.08 ± 1.09** | **97.52 ± 0.31** | **72.10 ± 1.74** | **96.54± 0.32** | 11.03 |

**Note.** Each row shows the number of output channels per stage in the encoder. Throughput is measured in iterations per second (it/s), with batch size = 64. Bold values indicate the best result for each metric.

classification performance. As the number of output channels increases, performance improves consistently across all classification metrics. The widest configuration, {1, 64, 128, 256, 512, 1024}, achieves the best overall results, reaching an accuracy of 84.08% ± 1.09%, a Top-3 accuracy of 97.52% ± 0.31%, a Macro F1 of 72.10% ± 1.74%, and a Macro AUC of 96.54% ± 0.32%. However, this improvement comes with a reduction in throughput, which decreases from 13.30 it/s in the narrowest setting to 11.03 it/s in the widest setting. The intermediate configuration, {1, 32, 64, 128, 256, 512}, already provides strong performance while maintaining slightly higher efficiency, suggesting a favorable trade-off between predictive performance and computational cost. Overall, the results indicate that increasing channel width improves representation capacity, although the gain is accompanied by lower throughput, as shown in Table 5.

*Normalization.* Table 6 compares two normalization strategies, `BatchNorm` and `LayerNorm`, while keeping the kernel-size configuration and channel widths fixed. The results show that `LayerNorm` consistently outperforms `BatchNorm` across all classification metrics, improving accuracy from 78.17% ± 2.12% to 82.54% ± 1.61%, Top-3 accuracy from 95.97% ± 0.60% to 97.01% ± 0.64%, Macro F1 from 64.11% ± 3.36% to 69.39% ± 3.13%, and Macro AUC from 94.35% ± 0.82% to 96.11% ± 0.49%. This improvement comes at a modest cost in efficiency, as throughput decreases from 12.91 it/s to 11.83 it/s. Overall, these results indicate that `LayerNorm` provides a more effective normalization strategy for our model (Table 6).

**Table 6**
Study on Normalization Strategy

| Metric | BatchNorm | LayerNorm |
|---|---|---|
| Accuracy (%) | 78.17 ± 2.12 | **82.54 ± 1.61** |
| Top-3 Acc (%) | 95.97 ± 0.60 | **97.01 ± 0.64** |
| Macro F1 (%) | 64.11 ± 3.36 | **69.39 ± 3.13** |
| Macro AUC (%) | 94.35 ± 0.82 | **96.11 ± 0.49** |
| Throughput (it/s) | **12.91** | 11.83 |

**Note.** Kernel size and channels are fixed. Bold values indicate the best result for each metric.

### *4.2. Comparison Across Other Works*

We compare `SpectraNet` against `DASH` (Muthukrishna et al. 2019), `ABC-SN` (Fortino et al. 2026). For `SpectraNet`, all spectra are processed using the preprocessing strategy described in Section 2, which is tailored to our architecture.

We perform a controlled comparison within the ZTF/SEDM-centered setting relevant to the present study. For `DASH` and `ABC-SN`, we adopt their published model architectures while retraining them using our data set, preprocessing pipeline, and cross-validation splits. Because these architectures were originally developed for different data characteristics and class definitions, this comparison should be interpreted as informative but not as a fully architecture-native or survey-independent benchmark. We therefore include the additional SNID-based comparison in

**Table 7**
Performance Comparison Across Different Models using Spectra from Multiple Epochs (or Phases) of the Transient in the ZTF/SEDM-centered Dataset used in This Study

| Metric (%) | ABC-SN | DASH | SpectraNet |
|---|---|---|---|
| Accuracy | 74.02 ± 1.34 | 65.23 ± 1.63 | **84.08 ± 1.09** |
| Top-3 Accuracy | 93.76 ± 0.70 | 89.40 ± 1.17 | **97.52 ± 0.31** |
| Macro F1 | 56.53 ± 1.72 | 48.09 ± 1.74 | **72.10 ± 1.74** |
| Macro AUC | 92.35 ± 0.27 | 88.22 ± 1.29 | **96.54 ± 0.32** |

**Note.** The comparison includes `ABC-SN`, `DASH`, and `SpectraNet`. All results are reported as the mean ± standard deviation across five cross-validation folds. Bold values indicate the best result for each metric.

**Table 8**
Per-class AUC Comparison Across Different Models using Multi-epoch Spectra in the ZTF/SEDM-centered Dataset When Available

| Class (AUC %) | ABC-SN | DASH | SpectraNet |
|---|---|---|---|
| SN Ia | 96.14 ± 0.36 | 92.63 ± 1.18 | **98.42 ± 0.24** |
| SN Ibc | 86.67 ± 1.10 | 77.94 ± 2.40 | **93.54 ± 0.85** |
| SN II | 91.69 ± 0.53 | 87.27 ± 1.26 | **96.03 ± 0.42** |
| AGN | 93.04 ± 1.30 | 91.99 ± 2.20 | **96.76 ± 0.97** |
| CV | 94.63 ± 1.40 | 91.11 ± 2.44 | **97.76 ± 0.66** |
| TDE | 91.94 ± 2.58 | 88.38 ± 2.41 | **96.76± 0.71** |

**Note.** Bold indicates the best result per class. The comparison is between `ABC-SN`, `DASH`, and `SpectraNet`. Values are reported as the mean over five-fold cross-validation.

Appendix A as a more directly comparable cross-method reference. All models are trained and evaluated on the same five-fold cross-validation protocol, and we report the mean ± standard deviation across the five folds for all evaluation metrics.

The overall performance comparison is summarized in Table 7, while the per-class AUC results are presented in Table 8.

## 5. Performance

The model is evaluated under both a primary-label configuration and an extended-label configuration with a larger number of classes. As the classification task becomes more fine-grained, performance degrades relative to the 6-class setting, reflecting the increased difficulty of distinguishing between closely related classes.

### *5.1. Performance on Primary Class Configuration*

The performance of the `SpectraNet` model under the 6-class configuration is shown in Figure 4. The confusion matrix exhibits a strong diagonal structure, indicating high classification accuracy across most classes. The best performance is achieved for SN Ia, while relatively lower accuracy is observed for TDE and AGN.

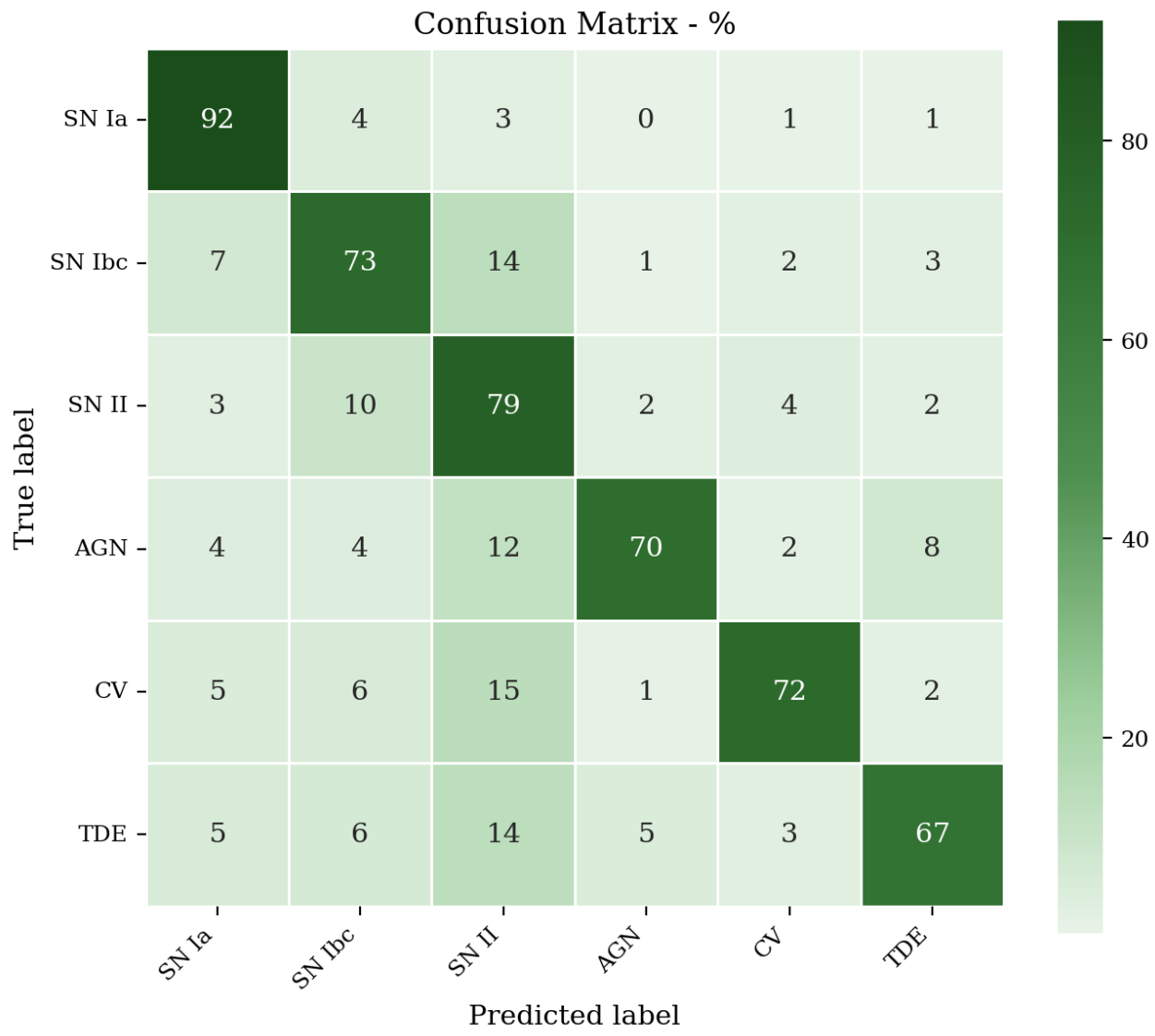

**Figure 4.** Confusion matrix of the `SpectraNet` model for the 6-class classification task shown as a row-normalized matrix. Each row sums to 100%, so the diagonal entries indicate class-wise completeness for the true labels.

Misclassifications are primarily concentrated among spectrally similar classes, particularly between SN Ibc and SN II, as well as between CV and SN II. This pattern likely reflects both intrinsic similarities in the underlying spectral features and limitations of the present model in separating closely related classes under the available data conditions.

### *5.2. Performance on Extended Class Configuration*

In this section, we evaluate the full `SpectraNet` model under a more detailed class configuration than that adopted in Table 3. This setting is designed to assess the generalization capability of the network in a more fine-grained classification task, including subclasses with fewer available samples (see Tables 1 and 2).

The confusion matrix in Figure 5 shows that the model retains a reasonable level of robustness under this extended class configuration, with a substantial fraction of predictions still concentrated along the diagonal. However, compared with the 6-class setting, the classification performance decreases for several subclasses. Misclassifications are more frequent among spectrally similar subclasses, particularly within the SN Ia and SN II groups, indicating that the finer-grained label definition introduces greater ambiguity in the classification task. The degradation is especially pronounced for subclasses with limited training samples, suggesting that class imbalance and reduced sample size further increase the difficulty of discrimination.

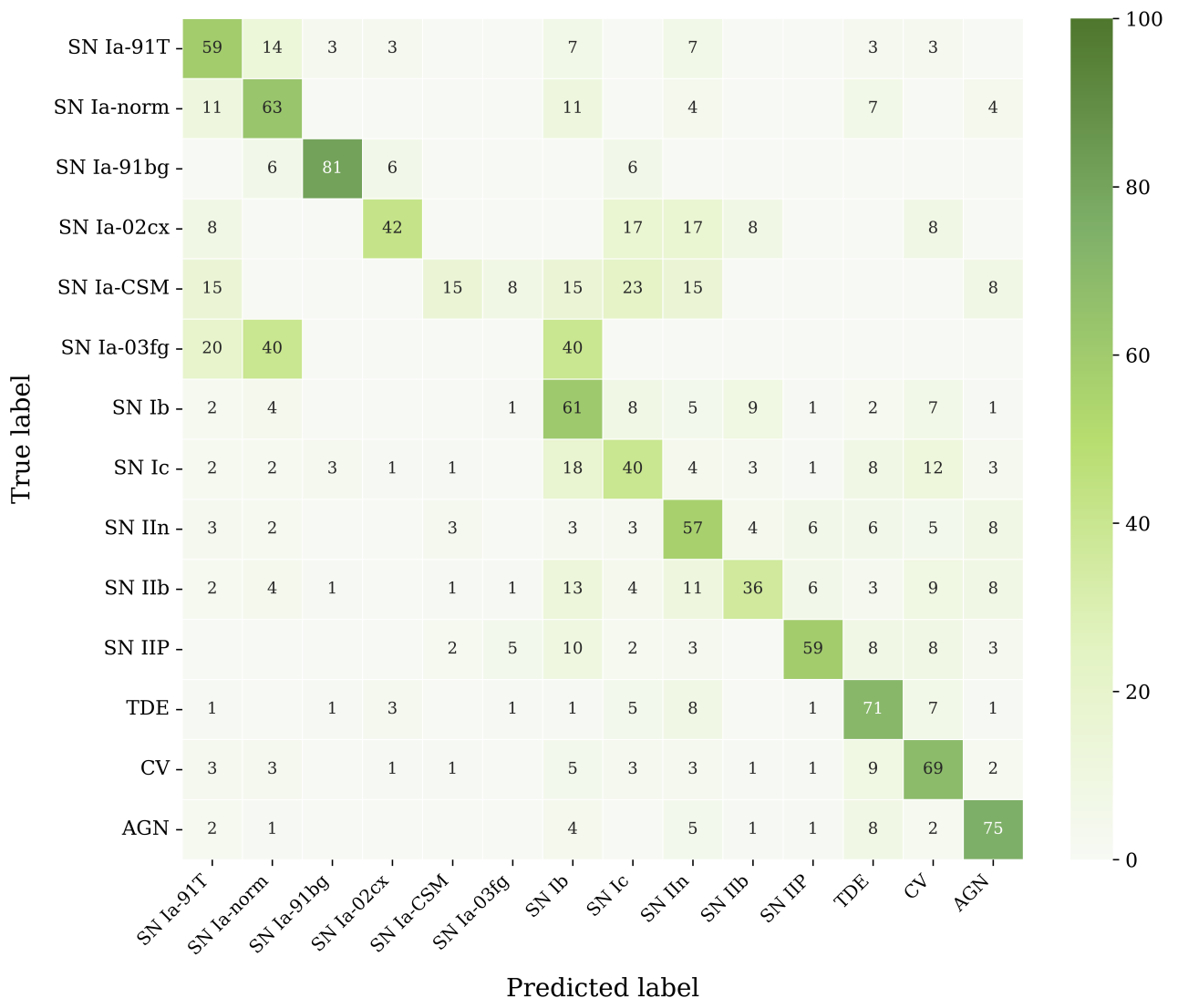

**Figure 5.** Confusion matrix of the full `SpectraNet` model under the extended class configuration shown as a row-normalized matrix. Each row sums to 100%, so the diagonal entries indicate class-wise completeness for the true labels.

## 6. Conclusion

Spectroscopy provides detailed information on composition, kinematics, redshift, temperature, and ionization of a source. Photometric data and/or images alone sometimes are not enough to provide an accurate classification of a source, and spectroscopy could reveal evolving absorption and emission features over varying timescales. `SpectraNet` has been developed in the framework of `AppleCiDEr` (Junell et al. 2026), which applies multi-modal learning to rapidly classify transients and potentially trigger spectroscopic follow-up.

`SpectraNet` is a 1D convolutional neural network designed for spectral classification. It utilizes a multi-branch convolutional design within each `SpectraBlock` to capture both narrow and broad features in astrophysical spectra. Each block contains three parallel convolutional paths with different kernel sizes (small, medium, large), followed by a $1 \times 1$ convolution to reduce the dimensionality and computational load. The network consists of five `SpectraBlock` units, an adaptive max pool, and two fully connected layers, working with preprocessed input data to classify spectra efficiently.

The network demonstrates strong performance across multiple evaluation metrics (e.g., accuracy, top-3 accuracy, and macro-averaged F1 score). It presents clear improvements over established baselines. Notably, the model performance on TDEs is particularly encouraging, as these rare events are often misclassified using traditional methods.

One of the most promising extensions of `SpectraNet` is the integration of multiple spectra from different evolutionary phases of the same transient within the network. Currently, the model accepts each spectral epoch independently, potentially ignoring evolution information that could significantly improve classification accuracy. On the primary 6-class task, `SpectraNet` achieves a macro F1 score of $72.10\% \pm 1.74\%$, outperforming the compared baselines. On the SNID-template data set discussed in Appendix A, our model achieves a macro F1 score of 0.8152, compared to 0.8076 for ABC-SN, indicating broadly comparable performance under this evaluation setting.

In conclusion, the `AppleCiDEr` pipeline adopts `SpectraNet` and presents a new automated transient classification framework for time-domain astronomy. As the field prepares for the Rubin Observatory Legacy Survey of Space and Time (LSST; Ivezić et al. 2019), such capabilities will become increasingly important for extracting scientific value from large-scale alert streams.

## Acknowledgments

The UMN authors acknowledge support from the National Science Foundation with grant Nos. PHY-2117997, PHY-2308862 and PHY-2409481.

N.R. is supported by DoE award #DE-SC0025599. Zwicky Transient Facility access for N.R. was supported by Northwestern University and the Center for Interdisciplinary Exploration and Research in Astrophysics (CIERA).

Based on observations obtained with the Samuel Oschin Telescope 48 inch and the 60 inch Telescope at the Palomar Observatory as part of the Zwicky Transient Facility project. ZTF is supported by the National Science Foundation under grants Nos. AST-1440341, AST-2034437, and currently Award #2407588. ZTF receives additional funding from the ZTF partnership. Current members include Caltech, USA; Caltech/IPAC, USA; University of Maryland, USA; University of California, Berkeley, USA; University of Wisconsin at Milwaukee, USA; Cornell University, USA; Drexel University, USA; University of North Carolina at Chapel Hill, USA; Institute of Science and Technology, Austria; National Central University, Taiwan, and OKC, University of Stockholm, Sweden. Operations are conducted by Caltech's Optical Observatory (COO), Caltech/IPAC, and the University of Washington at Seattle, USA.

SED Machine is based upon work supported by the National Science Foundation under grant No. 1106171.

The Gordon and Betty Moore Foundation, through both the Data-Driven Investigator Program and a dedicated grant, provided critical funding for SkyPortal.

We thank the anonymous referee for their valuable comments and suggestions, which helped improve the quality of this manuscript.

## Appendix
## Generalization on an Additional Spectral Dataset

Because our primary data set is tailored to early-time supernova classification, it differs from the SNID-based supernova template data sets commonly adopted in prior work. To enable a more direct comparison with previous studies and to further assess the generalization capability of our model, we additionally evaluate `SpectraNet` on the open-source SNID-template (Blondin & Tonry 2007) data set released by `ABC-SN`. This data set follows a two-fold cross-validation protocol. To make it compatible with our framework, we resample the spectra to match the input resolution required by our network, and then train and evaluate the model on this data set. The resulting confusion matrices are shown in Figures 6 and 7. Because the data set has already undergone a series of rigorous preprocessing steps, the amount of additional information that the model can learn is limited, leading to only minor performance differences between models. Under this setting, `SpectraNet` and `ABC-SN` show broadly comparable performance, with macro F1 scores of 0.8152 and 0.8076, respectively.

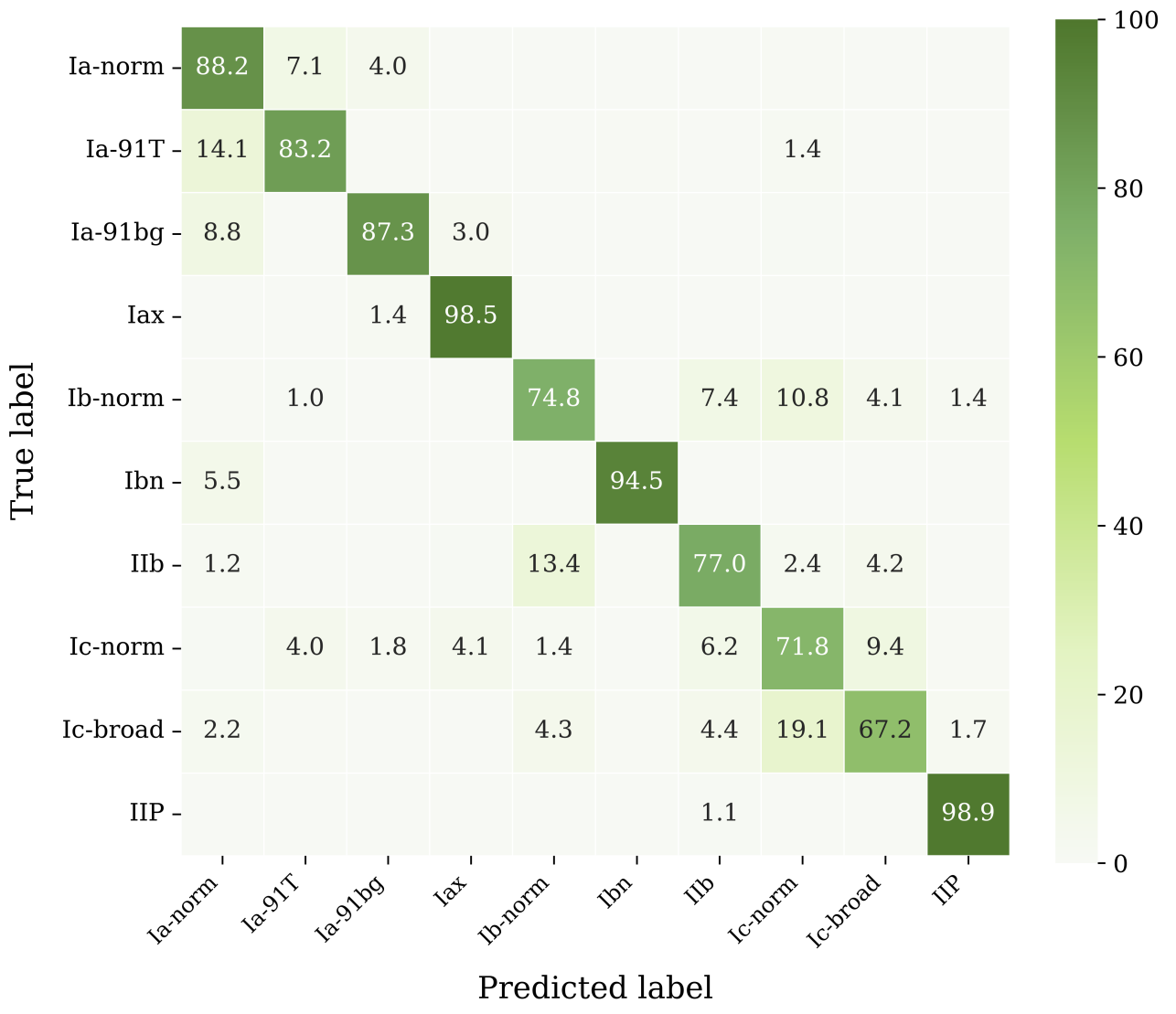

**Figure 6.** Confusion matrix of our model evaluated on the `ABC-SN` open-source SNID-template spectral data set shown as a row-normalized matrix. Each row sums to 100%, so the diagonal entries indicate class-wise completeness for the true labels.

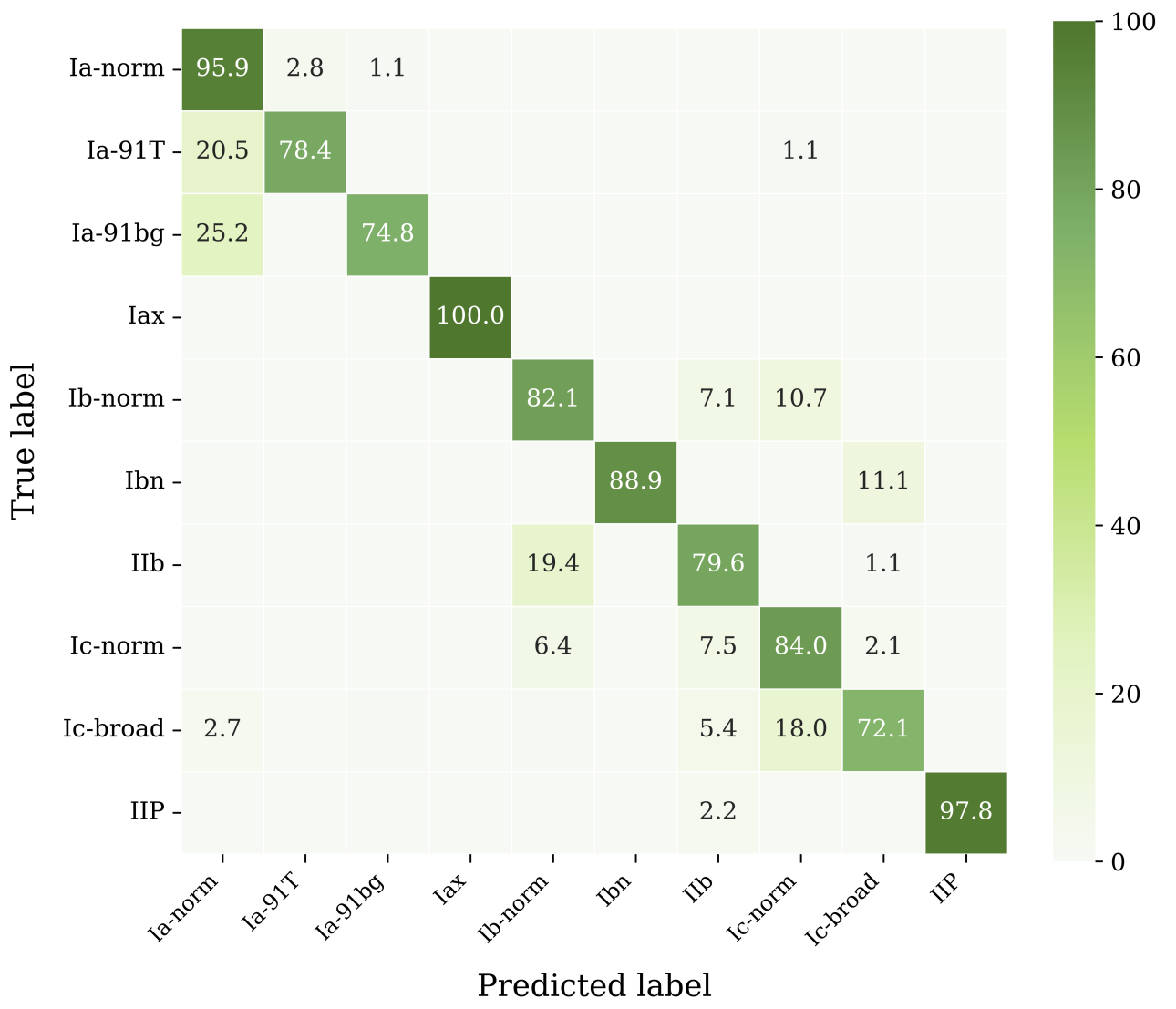

**Figure 7.** Confusion matrix of the `ABC-SN` model evaluated on the open-source SNID-template spectral data set shown as a row-normalized matrix. Each row sums to 100%, so the diagonal entries indicate class-wise completeness for the true labels.

## ORCID iDs

Argyro Sasli https://orcid.org/0000-0001-7357-0889
Alexandra Junell https://orcid.org/0000-0002-9380-7983
Felipe Fontinele Nunes https://orcid.org/0000-0001-7129-1325
Yu-Jing Qin https://orcid.org/0000-0003-3658-6026
Christoffer Fremling https://orcid.org/0000-0002-4223-103X
Sam Rose https://orcid.org/0000-0003-4725-4481
Theophile Jegou Du Laz https://orcid.org/0009-0003-6181-4526
Antoine Le Calloch https://orcid.org/0009-0009-7000-8343
Sushant Sharma Chaudhary https://orcid.org/0000-0003-1314-4241
Avyukt Raghuvanshi https://orcid.org/0009-0005-9422-0091
Nabeel Rehemtulla https://orcid.org/0000-0002-5683-2389
Jesper Sollerman https://orcid.org/0000-0003-1546-6615
Yashvi Sharma https://orcid.org/0000-0003-4531-1745
Judy Adler https://orcid.org/0009-0006-7265-2747
Tracy X. Chen https://orcid.org/0000-0001-9152-6224

Richard Dekany https://orcid.org/0000-0002-5884-7867
Reed Riddle https://orcid.org/0000-0002-0387-370X
Mansi M. Kasliwal https://orcid.org/0000-0002-5619-4938
Matthew J. Graham https://orcid.org/0000-0002-3168-0139
Michael W. Coughlin https://orcid.org/0000-0002-8262-2924